\documentclass[12pt,a4paper]{article}

  \usepackage{a4wide}
  \usepackage{latexsym}
  \usepackage{epsf}
  \usepackage{amssymb}
  \usepackage{graphicx}
  \usepackage{amsmath, cite}
  \usepackage{slashed,epsfig}
  \usepackage{bbm}
  \usepackage{bbm}
  \usepackage{amsmath,amssymb,amsthm}
  \usepackage{here}
\renewcommand{\d}{\textrm{d}}
\newcommand{\Real}{\textrm{I\!R}}

\newcommand{\e}{\textrm{e}}
\renewcommand{\d}{\textrm{d}}

\newcommand{\SU}{\mathop{\rm SU}}
\newcommand{\SO}{\mathop{\rm SO}}

\begin{document}

\begin{flushright}
\small UUITP-18/09\\
MAD-TH-09-06

\date \\
\normalsize
\end{flushright}

\begin{center}

\vspace{.5cm}

{\LARGE \bf{Towards
Classical de Sitter Solutions\\ \vspace{0.4cm}
in String Theory}}\\

\vspace{1.5 cm} {\large Ulf H.~Danielsson$^{1}$, Sheikh Shajidul Haque$^{2}$, Gary Shiu$^{2,3,4}$\\\vspace{0.2cm} and Thomas Van Riet$^{1}$ }\\
\vspace{0.6 cm} {\slshape $^1$ Institutionen f\"or fysik och
astronomi\\
Uppsala Universitet, Box 803, SE-751 08 Uppsala, Sweden}\\
\vspace{0.3 cm} {\slshape $^2$ Department of Physics, University of Wisconsin, Madison\\
Madison, WI 53706, USA}\\
\vspace{0.3 cm} {\slshape $^3$ School of Natural Sciences, Institute for Advanced Study\\
Princeton, NJ 08540, USA}
\\
\vspace{0.3 cm} {\slshape $^4$ Institute for Advanced Study, Hong Kong University of Science and Technology \\
Hong Kong, People's Republic of China}
\\
\vspace{0.4cm} {\upshape\ttfamily  ulf.danielsson,
thomas.vanriet@fysast.uu.se} \\
\vspace{0.1cm} {\upshape\ttfamily  haque@wisc.edu,
shiu@physics.wisc.edu}

\vspace{1.5cm}

{\bf Abstract}
\end{center}

\begin{quotation}
\small We investigate the type II string effective potential at
tree-level and derive necessary ingredients for having de Sitter
solutions in orientifold models
with fluxes. 
Furthermore, we examine some explicit O6 compactifications in IIA
supergravity on manifolds with $\SU(3)$-structure in the limit where
the orientifold sources are smeared. In particular, we use a simple
ten-dimensional Ansatz for four-dimensional de Sitter solutions and
find the explicit criteria in terms of the torsion classes such that
these de Sitter solutions solve the equations of motion. We have
verified these torsion conditions for the cosets and the Iwasawa
manifold and it turns out that the conditions cannot be fulfilled
for these spaces. However this investigation allows us to find new
non-supersymmetric AdS solutions for some cosets. It remains an open
question whether there exist $\SU(3)$-structure manifolds that
satisfy the conditions on the torsion classes for the simple de
Sitter solutions to exist.

\end{quotation}

\newpage

\pagestyle{plain}

 \tableofcontents

\section{Introduction}

The richness of string theory also presents us with a huge vacuum
degeneracy problem.  In lack of a dynamical principle to select a
unique vacuum, there are two broad approaches one typically takes.
In a bottom-up approach, one aims to construct models which realize
as many known properties of our universe as possible. The rationale
is that the criteria for a ``realistic" solution may significantly
reduce the space of vacua, and thus one can zero-in to a promising
subset which hopefully points to {\it the} vacuum that describes our
universe. Alternatively, one can quantify the ``likelihood" of our
universe by sampling the statistics of a vast number of vacua
without imposing the prior that such vacua resemble the one in which
we live. This latter approach is what underlines the idea of a
string landscape, and has often been invoked to address the
cosmological constant problem in string theory.

In contrast, the bottom-up approach has mainly been focussed on
particle physics aspects without much reference to the cosmological
constant. This is most apparent in local D-brane (and F-theory)
model building where the requirement of having the low energy
spectrum and interactions of the Standard Model or Grand Unified
Theories puts non-trivial restrictions on {\it local} properties of
the compactification, such as the types of singularities supported
in the internal space. These bottom-up constraints are powerful in
that they hold  for a large class of models without having to fully
specify the compactification details. Of course, they are only
necessary conditions, as {\it global} constraints such as moduli
stabilization and flux quantization can only be fully imposed with a
specification of UV completion. Nevertheless, they serve as a useful
guide in the search for realistic vacua before a complete model is
explicitly constructed.

Given the cosmological constant problem is a question that arises
only in the context of quantum gravity, it should play an equally
(if not more) important role in the selection of string vacua. A
natural question is whether there are analogous bottom-up
constraints on the underlying compactification in order for the
resulting string theory solutions to have positive 4D energy
density\footnote{A conventional wisdom is to search for realistic
vacua that preserve supersymmetry at the compactification scale, and
that supersymmetry is dynamically broken (e.g., due to strong
dynamics in the hidden sector) in the effective theory at lower
energies. However, not all realistic features
of the models (such as masses and couplings) necessarily
 persist after supersymmetry breaking and vacuum
uplifting.}. Naively, the answer is no since the cosmological
constant is defined only after all moduli are stabilized and so
details of compactifications are needed before this question can be
addressed. As we shall see, however, under some assumptions which
will be elaborated further, one can obtain a set of constraints on
the internal manifold valid for a large class of models without
specifying the compactification details. Our results thus suggest a
different strategy to search for de Sitter solutions, allowing us to
focus on promising regions of the landscape instead of constructing
them in a model by model basis.

Our investigation is guided by various no-go theorems, some appeared
in the recent literature \cite{Hertzberg:2007wc, Flauger:2008ad,
Caviezel:2008ik, Caviezel:2008tf, Haque:2008jz} and some we proved
along the way. We center our discussions  on Type II string theories
and their effective supergravity action since moduli stabilization
is more developed in the Type II duality frames. In particular, it
is well known by now that classical ingredients such as background
fluxes have the effect of fixing moduli \cite{Dasgupta:1999ss,
Greene:2000gh, Becker:2000rz, Becker:1996gj, Giddings:2001yu} (see
e.g. \cite{Douglas:2006es, Denef:2007pq, Blumenhagen:2006ci,
Grana:2005jc, Silverstein:2004id} for reviews). Although
non-perturbative effects are often invoked in scenarios of moduli
stabilization (e.g., in the Type IIB context of
\cite{Kachru:2003aw}), the full moduli dependence of such effects is
extremely difficult to determine explicitly. Therefore, much of the
work on the subject amounts to demonstrating (by zero mode counting)
that certain instanton effects crucial for moduli stabilization are
non-vanishing, rather than providing an explicit computation of
their magnitude and moduli dependence.

For ease of making our statements precise, we thus focus on finding
de Sitter solutions with only {\it classical} objects such as
fluxes, orientifold planes, and curvature along the lines of
\cite{Silverstein:2007ac, Caviezel:2008tf, Haque:2008jz,
Hertzberg:2007wc, Saltman:2004jh}, since their contributions to the
4D potential are explicitly computable. In the ``minimalist" spirit
of \cite{Haque:2008jz}, we do not consider introducing D-branes or
orbifolding the internal manifold even though these ingredients also
lead to a computable potential. This is because their presence also
implies new moduli such as those arising from open strings and
twisted sectors. In \cite{Hertzberg:2007wc, Silverstein:2007ac} it
has been argued that KK monopoles and NS5 branes lead to
contributions in the 4D effective potential that can enhance the
existence of de Sitter critical points.  However, since our ultimate
goal is to construct de Sitter solutions from a 10D point of view we
refrain from introducing these objects since it is far from clear
how the backreaction of such objects can be taken into account as to
have a reliable 4D de Sitter solution. Of course there are still
backreaction issues when one restricts to orientifolds, and
admittedly we have only been able to find solutions in the smeared
limit. It is nonetheless more likely that for configurations with
just orientifolds the backreaction can be computed and one would be
able to tell whether the de Sitter solution still exists.

Within the framework of this ``minimalist'' approach there appeared
some recent works on dS solutions in IIA \cite{Caviezel:2008tf,
Haque:2008jz}\footnote{For literature on non-classical dS solutions
in IIA we refer to \cite{Palti:2008mg,
Saueressig:2005es,Kallosh:2006fm}.}. It is one of our aims to improve
on these works since the proposed stable dS solution in
\cite{Haque:2008jz} turns out to not solve the 10D
equations of motion whereas the candidate example in \cite{Caviezel:2008tf}
is perturbatively unstable. Furthermore, because of the
complexity of the solution in \cite{Caviezel:2008tf} it is hard to
check that it really solves the 10D equations of motion\footnote{In
the sourceless case, that admits no dS solutions, there
exist arguments showing that the dimensional reduction is consistent
\cite{Cassani:2009ck}. }.

We investigate the effective potential for such Type II
compactifications and search for de Sitter critical points in models
with orientifold sources and fluxes on a compact internal manifold.
Our treatments for Type IIA and IIB theories are completely parallel
except for some obvious changes as one goes between these duality
frames. We derive several no-go conditions for the existence of de
Sitter solutions, and explore some explicit models that circumvent
them. In the specific case of $\SU(3)$-structure manifolds in IIA
with smeared O6 planes, we find de Sitter solutions that solve the
10D equations of motion when certain conditions on the torsion
classes are satisfied, even though the stability of such de Sitter
solutions needs to be checked once specific models are found.  On
the other hand, we verify that these torsion conditions are {\it
not} satisfied for the coset geometries. These examples illustrate
the utility and power of the no-go constraints. It remains an open
problem whether there exist $\SU(3)$-manifolds that satisfy the
conditions on the torsion classes for these simple de Sitter
solutions to be realized.

As an interesting aside we find that our analysis allows us to
construct new non-supersymmetric AdS solutions for some coset
geometries.

\section{The coupling and volume dependence of $V_{tree}$}
The number of scalar fields appearing in an effective 4D theory
after compactification depends on the specifications of the
compactification under consideration. Nonetheless there are 2
universal moduli that always appear, these are the string coupling
$\phi$ and the internal volume $\mathcal{V}$. The appearance in the
effective potential at tree-level is also universal, see for
instance \cite{Silverstein:2004id, Hertzberg:2007wc}. In the
following we re-derive these potential terms from type II
supergravity since we will need these to derive our nogo theorems in
the next section.

The metric Ansatz, in 10 dimensional string frame, that describes an
unwarped reduction to $3+1$ dimensions is
\begin{equation}
\d s_{10}^2 = \tau^{-2} \d s_4^2 + \rho\,\d s_{6}^2\,,
\end{equation}
where we have to take
\begin{equation}
\tau\equiv \rho^{3/2}\e^{-\phi}\,,
\end{equation}
in order to find 4D Einstein frame\footnote{In our conventions, the
10D string frame action is $ \int\sqrt{\mid g\mid }\e^{-2\phi}(\mathcal{R} +
4(\partial\phi)^2+\ldots)$.}.

The NSNS fluxes are the $H$ field strength and the metric flux. By
metric flux we mean that the internal manifold has non-zero Ricci
scalar. The energy contributions are
\begin{equation}
V_{R}= U_R\rho^{-1}\tau^{-2}\,,\qquad V_{H}=
U_{H}\rho^{-3}\tau^{-2}\,,
\end{equation}
where $U_H$ denotes the integrated flux, $U_H=\int_6 H^2$, and $U_R$
denotes minus the integrated curvature, $U_R=-\int_6\mathcal{R}_6$ .
We will use similar notation in the following that we consider
self-explanatory. For the RR $q$-form
fluxes we find
\begin{equation}
V_{q}^{RR}= U_q\tau^{-4}\rho^{3-q}\,.
\end{equation}
For D$p$ and O$p$ sources, with tension $T_p$, that fill the lower
4D spacetime and wrap a $(p-3)$-dimensional submanifold $\Sigma$ we
find
\begin{equation}
V_{Dp/Op}= \pm |\mu_p| Vol(\Sigma)\tau^{-3}\rho^{\frac{p-6}{2}}\,.
\end{equation}
The plus sign is for D-branes and the minus sign for O-planes.

From the above discussion we find that the form of the string
effective potential in $D=4$ at tree-level can be written as
\begin{equation}\label{potential}
\boxed{V_{tree} = a(\varphi)\tau^{-2} - b(\varphi)\tau^{-3} +
c(\varphi)\tau^{-4}\,,}
\end{equation}
where $\varphi$ denotes all scalars different from $\tau$ (including
$\rho$).


In the case the internal space is unwarped and compact one easily
verifies that the effective potential approach is correct  since the
$\partial_{\rho}V=0=\partial_{\tau}V$ equations correspond to
specific linear combinations of the 10D dilaton equation of motion
and the trace over the internal indices of the 10D Einstein
equations as shown in appendix B\footnote{When warping is present
one needs to be more careful in reducing the action. For instance,
there exist models that allow de Sitter solutions without sources
\cite{Gibbons:2001wy} (but with non-compact internal space),
although the $\rho, \tau$ appearance in the naively reduced scalar
potential would not allow for it.\\}.

\section{No-go theorems and minimal ingredients }
In this section we consider all orientifold compactifications and
focus on the form of the tree-level scalar potential. Since we
require the O-planes to fill 4D space and wrap some internal
submanifold, the O$p$-planes we consider have $p\geq 3$. If we
furthermore insist that the oientifolds do not break supersymmetry
explicitly so that the resulting dS solutions correspond to
supersymmetry breaking states in a supersymmetric theory, their
dimensionality should differ by a multiple of 4.
Finally, the O9-plane tadpoles are canceled by D9-branes which
introduce open string moduli. With the minimalist approach we pursue
here, we shall not consider this possibility though we expect our
considerations can be applied to the O9 cases as well. Therefore, we
end up with the following options in IIA: O4, O6, O8 and O4/O8 and
in IIB:  O3, O5, O7 and O3/O7.

\subsection*{The minimal ingredients}
As originally discussed in \cite{Silverstein:2007ac}, searching for
de Sitter critical points with small vacuum energy of the potential
corresponds to finding critical points of the quantity $4ac/b^2
\approx 1$ as a function of the other moduli $\varphi$:
\begin{equation}
\partial_\varphi\frac{4ac}{b^2}=0\,,\qquad \& \qquad \frac{4ac}{b^2}\approx
1\,.
\end{equation}
With this simple result we can easily construct no-go conditions for
dS solutions by investigating when $4ac/b^2$ allows for critical
points. By focussing on just the $\rho$-dependence of $4ac/b^2$ we
can give conditions that hold independently of the geometry and
other ingredients specific to a model. In particular we will list
the minimal ingredients that are necessary to have a critical point
of $4ac/b^2$ for type IIA/B supergravity with sources.

For the single type O$p$ reductions we have
\begin{equation}
\frac{4ac}{b^2}=\frac{\sum_q
U_q\rho^{6-p-q}}{Vol_{\Sigma}^2T_p^2}\,\,\Bigl(U_R\rho^2 +
U_H\Bigr)\,.
\end{equation}
The demand that $4ac/b^2$ is stabilised close to 1 shows that
\begin{equation}\label{closeto1}
U_R\rho^2 + U_H >0.
\end{equation}
From $\partial_{\rho} (4ac/b^2)=0 $ we deduce that
\begin{equation}\label{drho}
2U_R\rho^{7-p}\sum_qU_q\rho^{-q}=-\Bigl(U_R\rho^2 +
U_H\Bigr)\,\,\sum_q(6-p-q)U_q\rho^{5-p-q}.
\end{equation}
This equation combined with (\ref{closeto1}) and the fact that $U_H
$ and $U_q$ are all positive implies that for $p>4$ we need to have
$U_R>0$ and hence we need negatively curved internal spaces. In general we
deduce the following conditions from (\ref{drho}):
\begin{itemize}
\item O3 planes: When $U_R=0$ we need at least $U_H, F_1$ and $F_5$.
When $U_R\neq 0$ more possibilities arise.
\item O4 planes: When $U_R=0$ we need at least $U_H$, $U_0$ and
$U_q$ with $q>2$. When $U_R \neq 0$ more possibilities arise.
\item O5 planes: We minimally need positive $U_R$, $U_1$ and some other field
strength turned on.
\item O6 planes: The minimal conditions which were derived previously in
\cite{Haque:2008jz} and are positive $U_R$, $U_0$ and some $U_q$
with $q>2$ (or positive $U_R$, $U_0$, $U_H$ with $U_q$ with $q>0$.).
\item O7 $\&$ O8 planes: We cannot stabilise $4ac/b^2$.
\end{itemize}

For the O4/O8 and O7/O3 setup the expressions for $\frac{4 ac}{b^2}$
are more lengthy but a close look at the expressions shows that:
\begin{itemize}
\item O4/O8: One needs at least $U_R$ and $U_2$, or $U_0$ and $U_H$.
\item O3/O7: One needs at least $U_R$ and $U_3$, or $U_1$ and $U_H$.
\end{itemize}

The above derivations use the dependence of the effective potential
on $\rho$ and $\tau$ which is equivalent to the 10D dilaton equation
and traced internal Einstein equation in the smeared limit, as
explained in appendix B. The traced external Einstein equation just
fixes the value of the 4D cosmological constant, and contains no new
information. But in some cases one is able to use some extra
equation to find an extra relation. This was done in GKP
\cite{Giddings:2001yu}, where the $F_5$ equation of motion (or
Bianchi identity) was used in the traced external Einstein equations
to find extra nogo conditions. Let us briefly repeat the outcome of
that result and furthermore drop the assumptions of
\cite{Giddings:2001yu} that the 4D space is Minkowski and that the
internal space is a warped Calabi-Yau.

The Ansatz for $F_5$ in \cite{Giddings:2001yu} is
 \begin{equation}\label{assumption}
F_5 =(1+\star)\d\alpha\wedge \epsilon_4\,,
\end{equation}
where $\alpha$ is some function on the internal manifold (that is
even under the O3 target space involution in case there is an O3
source). The warped metric is given by
\begin{equation}
\d s_{10}^2 = \tau^{-2}\e^{2 A(y)}g^4_{\mu\nu}\d x^{\mu}\d x^{\nu}
+\rho\e^{-2 A(y)}g^6_{ij}\d y^i\d y^j\,,
\end{equation}
Repeating the same steps as in \cite{Giddings:2001yu} for O3 and O7
sources, one finds from the traced external Einstein equation and
the $F_5$ Bianchi identity the following condition
\begin{equation} \label{pseudoBPS}
\Box (\e^{4A}-\alpha)=\mathcal{R}_4 + \frac{\e^{2A}}{6Im\tau}
|iG_3-\star_6 G_3|^2 +\e^{-6A}|\partial(\e^{4A}-\alpha)|^2\,.
\end{equation}
If we integrate the equation on both sides over the internal
manifold then we clearly find that $\mathcal{R}_4>0$ is impossible
since the other 2 terms on the right hand are manifestly
non-negative. This excludes any dS vacuum given the assumption for
the $F_5$ field strength (11).

Let us therefore examine this assumption (\ref{assumption}). Clearly
for Calabi Yau spaces this assumption is necessary since there exist
no non-trivial 1- or 5-cycles. But here we drop the Calabi-Yau
assumption, such that one can in principle have
\begin{equation}\label{assumption2}
F_5 =(1+\star)A \wedge \epsilon_4\,,
\end{equation}
where $A$ is some cohomoligical non-trivial one-form. In this case
one cannot derive equation (\ref{pseudoBPS}) to exclude dS
solutions. However for O3 planes (\ref{assumption2}) is excluded
since a non-trivial one-form would be projected out by the O3
involution. Hence the GKP argument also applies here and
demonstrates that the minimal ingredients derived above are not
sufficient since there do not exist tree-level dS solutions. This
leaves O5 models as the only possibilities (since we already
excluded O7).

\subsection*{``Pure flux'' models}
In this subsection we check whether the minimal ingredients can be
satisfied in the simplified situation \emph{that cycles thread by
the field strengths and the cycles wrapped by the sources are closed
but non-exact.}

Consider a field strength $F_p=\d C_{p-1}$. When we truncate all 4D
vectors and 4D tensors, the dimensional reduction is
\begin{equation}
\hat{C}_q=\chi_i\Lambda^i_q\,,\qquad \hat{F}_p=\d \hat{C}_{p-1} +
\Sigma_p\,,\qquad
\end{equation}
where $\Sigma_p$ are non-trivial elements of the $p$-th cohomology
class $\Omega^P(M,\Real)$ of the internal manifold $M$. The $\chi^i$
are 4D scalar fields (the gauge potential moduli) and the
$\Lambda^i$ are a set of $p$-forms on $M$, chosen such that the
reduction corresponds to a consistent truncation. We define ``pure
flux'' solutions as solutions for which we truncate all the gauge
potential moduli: $\chi^i=0$.

One needs to take into account the orientifold involutions to
understand what kind of fluxes are allowed by the orientifolds. An
orientifold action is a combination of different involutions. There
is always a target space involution $\sigma$ and the world-sheet
parity operation $\Omega$, exchanging left and right movers. The
fixed point set of the geometric involution $\sigma$ defines the
position of the orientifold. In some case one needs to add the
involution $(-1)^{F_L}$, with $F_L$ the left-moving fermion number.
We have the following transformation properties:
\begin{align}
\Omega: & \quad + \quad \bigl\{\phi, g, C_1, C_2\bigr\}\,,\qquad  -\quad \bigl\{C_0, B_2,C_3,C_4\bigr\}\,, \\
(-1)^{F_L}: & \quad + \quad \bigl\{ \phi, g,
B_2\bigr\}\,,\,\,\,\quad\qquad -\quad \bigl\{C_0, C_1, C_2, C_3,
C_4\bigr\} \,.
\end{align}
%
It can  be shown that, in order to divide out by symmetries of the
string theory, the orientifolds come with the following actions:
\begin{align}
& (-)^{F_L}\Omega\sigma\qquad : \qquad O3, O4, O6, O7, O8\,,\nonumber\\
&\Omega\sigma\qquad\qquad\,\,\,\, : \qquad O5,
O9\,.\label{worldsheetinv}
\end{align}
Hence to understand which degrees of freedom and which fluxes are
allowed by the orientifold one multiplies the worldsheet involutions
(\ref{worldsheetinv}) for a certain field $C$ (or flux $F$) and one
considers how many legs of $C$ (or flux $F$) are in the orientifold
direction and how many are transversal. The latter is necessary to
check the parity of the field, or flux, under $\sigma$. The total
product should be even. Let us investigate this for the O4, O5 and
O6 cases.\\

$\bullet $\underline{The O4 model}

The Bianchi identity
\begin{equation}
\d F_4= H\wedge F_2 + \delta(O4)\,,
\end{equation}
turns out problematic: $F_2$ has to thread a cycle with one leg in
the O4 and another leg outside. If this flux is wedged with $H$ we
have a 5-form with at least one leg inside of the O4. This 5-form is
hence of a different type then the 5-form distribution
$\delta(O4)$, which has all legs outside of the O4 plane. \\

$\bullet$ \underline{The O5 model}

The Bianchi identities for $F_3$ and $F_5$
\begin{equation}
\d F_5 =H\wedge F_3\,,\qquad \d F_3= F_1\wedge H +\delta(O5)\,,
\end{equation}
where $\delta(O5)$ is a form distribution with 4 legs in the space
transversal to the O5 plane. To evaluate these constraints we have
to keep in mind that $F_1, H$ and $F_5$ are odd under the O5
worldsheet operation involution and $F_3$ is even. Hence the $F_1,
H$ fluxes point in the transversal directions and the $F_5$ flux
should have an odd number of legs along the transversal directions,
whereas the $F_3$ should have an even number.\\

$\bullet$ \underline{The O6 model}

The $F_2$ Bianchi identity
\begin{equation}
\d F_2=mH +\delta(O6)\,.
\end{equation}
demonstrates that $H$ is needed to cancel the tadpole. Since $H$ needs to thread a cycle
transversal to the O6 it is the same form type as the form
distribution of the O6 source, such that it can indeed cancel the
tadpole. This is an attractive feature of these models.\\
%

Let us consider some examples. In case the internal space is a
direct product of two 3-dimensional spaces $\mathcal{M}_3$ there is
a straightforward way to define the O6 target space involution
$\sigma$:
\begin{equation}
\sigma: \qquad(y_1, y_2, y_3, \bar{y}_1, \bar{y}_2,
\bar{y_3})\leftrightarrow (\bar{y}_1, \bar{y}_2, \bar{y_3}, y_1,
y_2, y_3)
\end{equation}
where the $y$ and $\bar{y}$ represent coordinates on the 3D spaces.
Then there is one O6 plane at the three-cycle spanned by the
3-surface, $y_i=\bar{y}_i$. Of course, there are other ways to
define O6 planes, but this one is exceptionally easy.

In reference \cite{Haque:2008jz} some examples were studied where
$\mathcal{M}_3$ are all 3D unimodular group manifolds and where
$\mathcal{M}_3$ is the Weeks manifold (a compactification of the
hyperboloid $\SO(3,1)/\SO(3)$). In the group manifold case it turned
out that the other metric moduli, typical to group manifolds, have a
runaway behavior in the $4ac/b^2$ expression, excluding any dS
solutions. The Weeks manifold on the other hand has no moduli apart
from $\rho$ and $\tau$. If we insist on not turning on massive shape
moduli (that could be runaway) the possible fluxes are $F_0$, $H$
and $F_6$ . When all these are turned on we have (ignoring all
numerical factors)
\begin{equation}
\frac{4ac}{b^2}\propto \rho^2 + \rho^{-2} + \rho^0 + \rho^{-6}
\end{equation}
and this shows that a dS can be found if one can tune the numerical
factors, as turns out to be the case \cite{Haque:2008jz}. However
this model fails to be a 10D solution since the $F_4$ equation of
motion is not satisfied when $F_6\neq 0$. If we put $F_6=0$ we loose
the dS solution, since
\begin{equation}
 \frac{4ac}{b^2}\propto \rho^2 + \rho^0\,.
\end{equation}
A similar problem seems present in the model of
\cite{Silverstein:2007ac} where there is also a non-zero $F_6$ flux.

It turns out that this happens more generically: one can find models
for which one can stabilise the $4ac/b^2$ quantity, but if one then
insists on satisfying all the 10D form equations of motion one finds
exactly the terms needed for a solution to be forbidden. We illustrate
this with one more example that captures the essentials. For that we
take $\mathcal{M}_3=\mathbb{H}_2\times S_1$, where $\mathbb{H}_2$ is
the compact 2D hyperboloid. The only modulus it has is the breathing
mode. The metric Ansatz then is
\begin{equation}
\d s_{10}^2 =\tau^{-2}\d s_4^2 +\rho\bigl(\frac{1}{\phi}\d
\mathbb{H}^2 +\phi^2\d y^2 + \frac{1}{\phi}\d \bar{\mathbb{H}}^2
+\phi^2\d\bar{y}^2\bigr)\,.
\end{equation}
So, there are 3 scalars, $\tau, \rho$ and $\phi$, where the latter
measures the relative size of the hyperboloid and the circle. The
cycles that can be thread with fluxes, taking into account the
parity of the O6 are:
\begin{align}
& H_3:\quad \epsilon_2\wedge \d y - \bar{\epsilon}_2\wedge \d \bar{y} \,,\\
& F_2:\quad \epsilon_2 - \bar{\epsilon}_2\,,\qquad \d y\wedge \d \bar{y}\,,\\
& F_4:\quad \epsilon_2\wedge\bar{\epsilon}_2\,,\qquad(\epsilon_2
-\bar{\epsilon}_2)\wedge\d y\wedge\d\bar{y}\,,
\end{align}
where $\epsilon_2$ is the volume element on $\mathbb{H}_2$. However
upon using the 10D form equations one finds that the two
$F_2$-fluxes need to vanish. The $F_4$ fluxes are not constrained
and the $H$-flux is inversely proportional to the Romans mass
$F_0$. If we ignore all numerical factors, we obtain the following
expression
\begin{equation}
\frac{4ac}{b^2}\propto \phi\rho^2 + (\phi^5+\phi^{-1})\rho^{-2} +
(\phi^4 +\phi^{-2})\rho^{-4}\,.
\end{equation}
It is not possible to stabilise $\rho$ and $\phi$ at the same time.
To see this clearly we make the following redefinition
$\phi=\rho^{-2}\phi'$ and find
\begin{equation}
\frac{4ac}{b^2}\propto \phi' +
(\phi'^5\rho^{-10}+\phi'^{-1}\rho^2)\rho^{-2} + (\phi'^4\rho^{-8}
+\phi'^{-2}\rho^4)\rho^{-4}\,.
\end{equation}
such that all powers in $\rho$ are negative. One can readily check
that when both $F_2$ fluxes are turned on, this problem disappears.
So, it is really the information contained in the 10D form equations
that spoil the putative dS solution.

\section{O6 models on $\SU(3)$-structure manifolds} In this
section we reverse our strategy. Instead of investigating the scalar
potential coming from a specific internal manifold with fluxes and
then imposing the tadpole conditions,
we consider a
whole class of internal manifolds with an Ansatz for the fluxes that
solves the 10D form equations from the outset\footnote{We stress that
it is not necessary to investigate the 10D equations as long as one
performs a consistent dimensional reduction, which we believe can be
done for the models under consideration. However, we have found it
easier to analyse the 10D equations instead of performing the
reduction.}.

Since fluxes backreact the internal spaces to generalised Calabi-Yau
spaces we take as a starting point a general class of
$\SU(3)$-structure manifolds defined by two torsion classes $W_1$
and $W_2$. Consider the canonical real two-form $J$ and the complex
three-form $\Omega=\Omega_R+i\Omega_I$ built out of the everywhere
non-vanishing spinor on the internal manifold. We have the following
characteristic equations
\begin{align}
&\d J = -\frac{3i}{2}W_1\Omega_R\,,\\
&\d\Omega = W_1 J\wedge J + W_2\wedge J\,,
\end{align}
where we assume that $W_1$ is an imaginary zero-form and $W_2$ is an
imaginary two-form.

These kind of $\SU(3)$-structure spaces have been shown to allow for
supersymmetric AdS$_4$ solutions \cite{Lust:2004ig, Koerber:2008rx,
Caviezel:2008ik} with and without sources. Below we generalise the
AdS Ansatz of \cite{Lust:2004ig, Koerber:2008rx, Caviezel:2008ik}
and check whether it can give rise to dS$_4$ solutions. For the
readers' convenience we added appendix C that contains our IIA
conventions and appendix D that contains useful formulae involving
$\SU(3)$-structures.

Our 10D Ansatz for the forms is
\begin{align}
&F_2=\e^{-3\phi/4}f_1 J + i\e^{-3\phi/4}f_2W_2\,,\\
&H=\e^{\phi/2}h\Omega_R\,,\qquad F_0=\e^{-5\phi/4} m\,,\\
&F_4=\e^{-\phi/4}g_1\epsilon_4 + \e^{-\phi/4}g_2J\wedge J\,.
\end{align}
Concerning the O6 plane source we assume the same as in
\cite{Caviezel:2008ik} that it is smeared and that it wraps the
calibrated submanifold dual to $\Omega_R$  such that the Bianchi
identity reads
\begin{equation}
\d F_2=mH+\mu\Omega_R\,,
\end{equation}
where in this convention positive $\mu$ implies net orientifold
charge. The 10D Bianchi and form equations are solved if the flux
parameters obey
\begin{align}
 g_1h &=-3ig_2W_1\,,\label{eq1}\\
 ihW_1&=2f_1g_2- g_1g_2 +\frac{1}{2}mf_1\,,\label{eq2}\\
 h &= 2f_2g_2 - mf_2\,,\label{eq3}\\
f_2\frac{|W_2|^2}{8}&=mh +\frac{3i}{2}f_1 W_1 +
\e^{3\phi/4}\mu\,,\label{eq4}
\end{align}
and the following form equation is satisfied\footnote{One can prove
that this assumption fixes the constant of proportionality to become
$\d W_2=-(i|W_2|^2/8)\, \Omega_R$.}
\begin{align}
\d W_2\propto \Omega_R\,.
\end{align}
From here on we just use the Bianchi identity (\ref{eq4}) to
determine the sign and the magnitude of $\mu$. Of course, in an explicit
model, the magnitude of the net orientifold charge cannot be chosen at
will, since (i) the orientifold plane charges, just like D-brane charges, are quantized, and
(ii) orientifold planes cannot be stacked like D-branes and their
number is fixed through the number of $\mathbb{Z}_2$ involutions on the internal
manifold.

For specific values of the flux parameters $f_1, f_2, h, g_1, g_2$
one obtains the supersymmetric AdS solutions of \cite{Lust:2004ig,
Caviezel:2008ik}. These solutions have
\begin{align}
& f_1=\frac{i}{4}W_1\,,\qquad f_2=1\,,\qquad
h=-\frac{2m}{5}\,,\label{lust1}\\
& g_1=9f_1\,,\qquad g_2=\frac{3m}{10}\,.\label{lust2}
\end{align}
However, these ingredients are also sufficient to evade the usual dS
no-go theorems. It is therefore interesting to understand whether
there are other non-supersymmetric solutions in the 5-dimensional
parameter-space ($f_1, f_2, h, g_1, g_2$).

The most constraining 10D equation is the internal Einstein equation
(\ref{Einstein eq}). For the manifolds under consideration there
exist explicit expressions for the Ricci tensor in terms of the
forms $J, \Omega, W_2$ \cite{bedulli-2007-4, Ali:2006gd}:
\begin{equation}
\mathcal{R}_{mn}=-\frac{3i}{4}(\Omega_R)_n^{\,\,ps}\partial_{[p}(W_2)_{sm]}
-\frac{1}{4}W_1(W_2)_{mr}J_{n}^{\,\,\,r} -
\frac{1}{2}(W_2)_{mq}(W_2)_{n}^{\,\,\,q}
+\frac{5}{4}g_{mn}|W_1|^2\,.
\end{equation}
This clean expression implies we can verify in all generality the
10D equations of motion. The main clue to solve the 10D Einstein
equations is the understanding of which tensors on both sides of the
equation are independent. Clearly the traceless parts have to be
equal. The problem divides into two cases
\begin{align}
\text{case}~1 &:\qquad (W^2_2)_{ij} \neq \frac{W_2^2}{6}g_{ij} + i\alpha (JW_2)_{ij}  \,,\\
\text{case}~2 &:\qquad (W^2_2)_{ij} = \frac{W_2^2}{6}g_{ij} +
i\alpha (JW_2)_{ij}\,,
\end{align}
with $\alpha$ some real number different from zero\footnote{ In case
2 one can also verify that $\alpha\neq 0$. To show this note that
$J$ and $W$ commute as matrices and therefore can be complex
diagonalised at the same time. Using this as a starting point one
finds that $W^2$ cannot be proportional to the metric when at the
same time keeping $JW$ traceless.}. Case 1 is the most general case
and leads to the most restrictions. In case 2 the Einstein equations
enforces less restrictive conditions and we will show that dS solutions are
possible in this case.

\subsection*{The non-degenerate case} Let us first discuss case 1
and demonstrate that the only solutions are the supersymmetric AdS
solutions constructed in \cite{Lust:2004ig, Koerber:2008rx,
Caviezel:2008ik}. If we just focus on the tensors different from
$g_{ij}$ in the internal Einstein equation we find two conditions
from equating the coefficients in front of the $W^2_{ij}$ and
$(JW)_{ij}$ tensors on both sides of the Einstein equation:
\begin{equation}
 f_2 =\pm 1\,,\qquad
 -\frac{1}{4}W_1=if_1f_2\,.
\end{equation}
Combined with the equations (\ref{eq1}-\ref{eq4}) we uniquely find
the known supersymmetric AdS solutions (\ref{lust1}, \ref{lust2}).
In fact, without analysing the Einstein equation, supersymmetry
would immediately lead to these values for the fluxes and susy would
guarantee that the Einstein and dilaton equations are solved.  Since
dS vacua are not supersymmetric there is more work in order to check
when there is a solution.\footnote{ However, recently it has been
shown that some non-susy vacua have the same integrability
properties as the susy vacua \cite{Lust:2008zd}. We did not pursue
this possibility further.}

\subsection*{The degenerate case with $F_4=0$}

Let us now consider case 2. The traceless part of the Einstein
equations now imposes just one condition
\begin{equation}\label{alpha}
(-f^2_2 +1)\alpha = -2f_1f_2 + \frac{i}{2}W_1\,.
\end{equation}

First we consider the simplified case where $F_4=0$. From here on we
leave $h$ and $m$ free and solve all quantities in terms of these
two flux numbers. Furthermore the ratio $h/m$ is important enough to
deserve a separate name
\begin{equation}
\beta=\frac{h}{m}\,.
\end{equation}
Then the equations (\ref{eq1}-\ref{eq3}) imply
\begin{equation}
f_1=2\beta i W_1\,,\qquad f_2=-\beta\,,
\end{equation}
The $F_2$ Bianchi identity (\ref{eq4}) leads to
\begin{equation}\label{bianchi2}
\frac{\e^{3\phi/4}\mu}{\beta m^2}=-1 -\frac{1}{m^2}(3|W_1|^2 +
\frac{1}{8}|W_2|^2)\,.
\end{equation}
From this we observe that without source we cannot have a solution
and that $\beta>0$ corresponds to net D6 charge and $\beta<0$ to net
O6 charge. In case we are interested in dS solutions we therefore
need $\beta<0$.

The remaining equations to verify are the traced internal Einstein
equation and the dilaton equation, which are equivalent to the
$\partial_{\tau}V=\partial_{\rho}V=0$ equations
\begin{align}
\partial_{\rho}V=0:\qquad&\qquad -V_R -3V_H + 3V_0 + V_2=0 \label{eq5} \,,\\
\partial_{\tau}V=0:\qquad&\qquad -2V_R -2V_H - 4V_0 -4 V_2
\label{eq6}
-3V_{O6/D6}=0\,,
\end{align}
where
\begin{align}
& V_R=-\frac{15}{2}|W_1|^2 +\frac{1}{4}|W_2|^2 \,, \qquad V_0=\frac{m^2}{2}\,,\qquad  V_H=2h^2\,,\label{potentials0}\\
& V_2=\frac{1}{4}\,(6f_1^2 +f_2^2|W_2|^2 )\,,\qquad
V_{O6}=-4\mu\e^{3\phi/4}\label{potentials1} \,.
\end{align}
In order to verify that we have a solution we must solve (\ref{eq5})
and (\ref{eq6}) for $|W_1|^2$ and $|W_2|^2$ and check when the
expressions are positive. The solutions are
\begin{align}
&|W_1|^2 =\frac{-m^2}{81\beta}\,\Bigl(5 +16 \beta - 20\beta^2 - 28\beta^3 \Bigr)\,,\\
&|W_2|^2 =\frac{-2m^2}{27\,\beta\,(\beta + 1)}\,\Bigl(25 +24\beta
 -56\beta^2 +192\beta^3 +112\beta^4\Bigr) \,.
\end{align}
Clearly both expressions are positive when $\beta$ is negative and
sufficiently close to zero. From the Bianchi identity we know that
this also implies a net orientifold charge. In order to know what
the sign of the 4D cosmological constant is one observes that
equations (\ref{eq5}) and (\ref{eq6}) imply (only when $F_4=0$)
\begin{equation}
V=\frac{2}{3}(V_0-V_H)\quad\Longrightarrow\quad V>0:\quad
\beta^2<\frac{1}{4}\,.
\end{equation}
Hence a small negative $\beta$ is nicely consistent with a de Sitter
solution! To understand what kind of solutions are possible we
present some plots. In figure 1 we plot $|W_{1}|^{2}$ and
$|W_{2}|^{2}$ as functions of $\beta$. A solution exists when both
expressions are positive. In figure 2 we plot $V$ and the two
mass$^{2}$ eigenvalues in the $\rho$ and $\tau$ directions as
functions of $\beta$. From the figures we see that a value of
$\beta$ between roughly $-2$ and $-1$ gives rise to a
non-supersymmetric AdS vacuum that is stable in the $\rho$ and
$\tau$ directions. We also note that we have dS vacua
with a tachyonic direction for small negative values of $\beta$.%
\begin{figure}[t]
\begin{center}
\includegraphics[trim=0.000000in 0.000000in -0.002081in 0.000000in, height=2.0617in,
width=3.1514in]{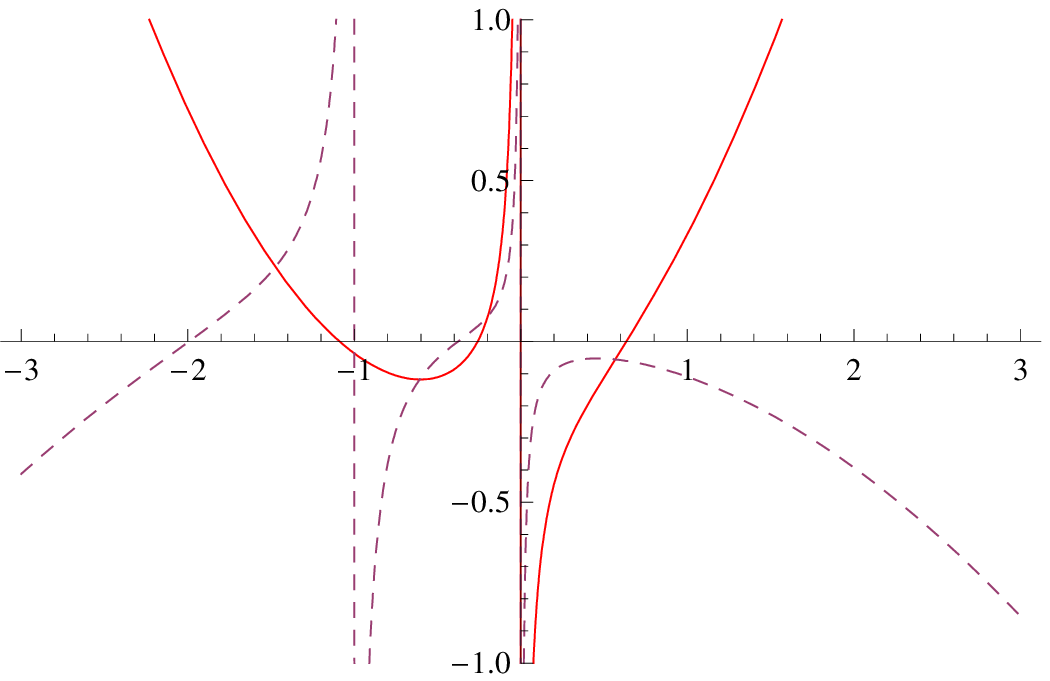} \caption{\small \emph{$\left\vert
W_{1}\right\vert ^{2}$ and $10^{-2}\times\left\vert W_{2}\right\vert
^{2}$ (dashed) as functions of $\beta$ when $F_4=0$.}}
\end{center}
\end{figure}


\

\ \
\begin{center}
\includegraphics[
height=2.0617in, width=3.1505in
]%
{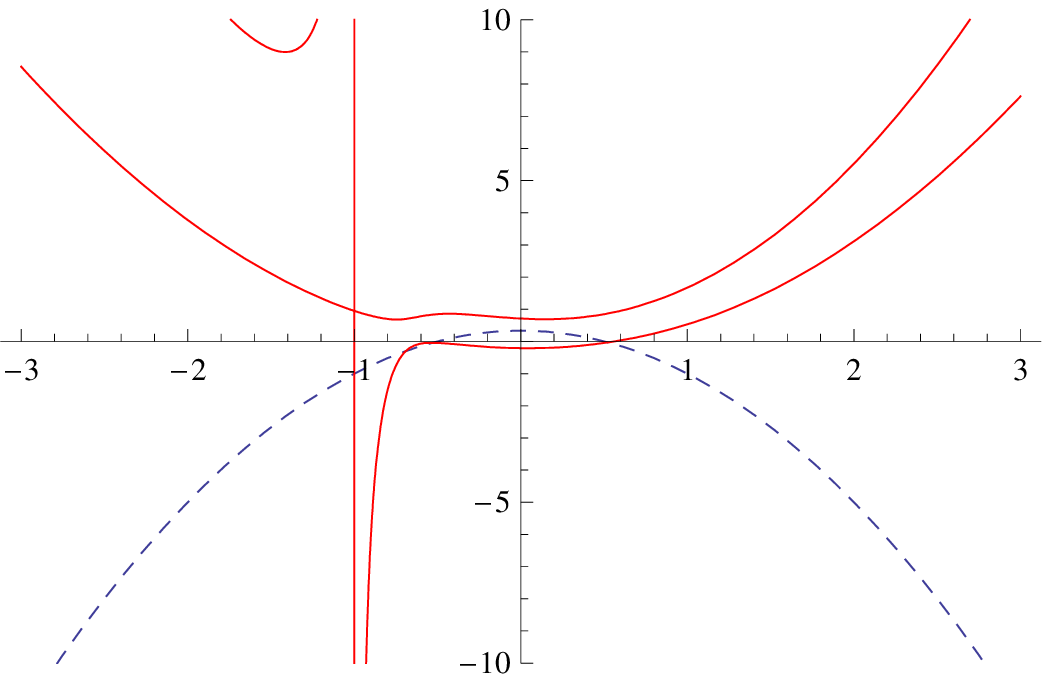}%
\\
Figure 2: {\small \emph{The two $10^{-1}\times$ mass$^{2}$
eigenvalues and $V$ (dashed)  as functions of $\beta$ when
$F_4=0$.}}
\label{Figure 2}%
\end{center}

\subsection*{The degenerate case with $F_4\neq 0$}
In what follows it is useful to also define a new fraction
\begin{equation}
\gamma\equiv\frac{g_2}{m}\,.
\end{equation}
We can solve $f_1, f_2$ and $g_2$ in terms of $\beta,\gamma, m$ and
the torsion classes as follows
\begin{align}
&f_2 = \frac{\beta}{2\gamma - 1}\,,\qquad g_1= \frac{-3\gamma i
W_1}{\beta}\,,\qquad f_1= \frac{\beta
-\frac{3\gamma^2}{\beta}}{\frac{1}{2}+2\gamma}iW_1\,.
\end{align}
Then the $F_2$ Bianchi identity (\ref{eq4}) is given by
\begin{equation}
\frac{\e^{3\phi/4}\mu}{\beta}=-m^2 + \frac{|W_2|^2}{16\gamma-8}
-\frac{\bigl(3-\frac{9\gamma^2}{\beta^2}\bigr)}{1+4\gamma}|W_1|^2
\,.
\end{equation}
An interesting effect of non-zero $\gamma$ is that the Bianchi
identity can be satisfied with zero source $\mu=0$. The
contributions to the potential are now ($V_R, V_{O6}, V_0, V_2$
remain unaltered)
\begin{equation}\label{potentials2}
V_4= 6g_2^2\,,\qquad V_6= \frac{1}{2}g_1^2\,.
\end{equation}

Having established this we can repeat the same kind of analysis as
above. One rewrites the $\partial_{\rho} V=\partial_{\tau} V=0$
equations in terms of $\beta ,\gamma, m, |W_1|^2 ,|W_2|^2$ and
checks when there exists solutions, i.e., when the solutions for
$|W_{1,2}|^2 $ in terms of $(\beta,\gamma, m)$ are
positive\footnote{It turns out that $m^2$ just sets the overall
scale and one can therefore just take $m^2=1$. Then one is left with
$\beta, \gamma$.}. Below we present plots of $|W_{1,2}|^2$ in terms
of $\beta$ for $\gamma=0.1$. From figure 3 and 4 we see that dS
solutions, stable in the $\rho,\tau$-directions exist for $\beta$
between about $-0.207$ and $-0.190$. Note that while a critical dS
is easy to achieve, there is just a tiny little window available for
a solution stable in the $\rho, \tau$-directions.
\begin{center}
\includegraphics[
height=2.0323in, width=3.1514in
]%
{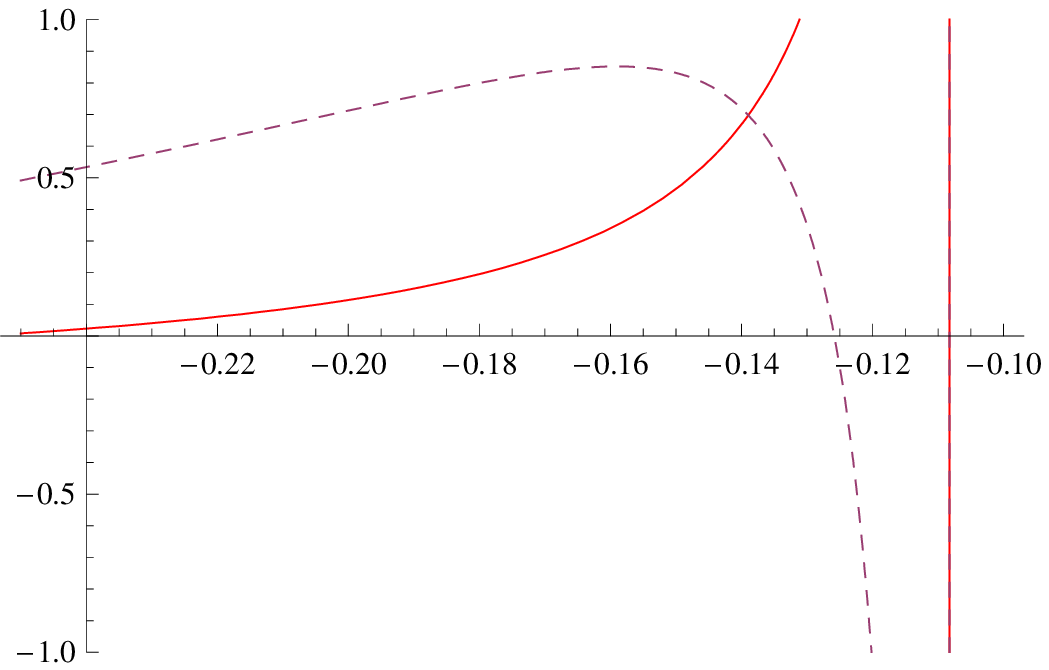}%
\\
{\small Figure 3:  \emph{$\left\vert W_{1} \right\vert ^{2}$ and
$10^{-1}\times\left\vert W_{2} \right\vert ^{2}$ (dashed) as
functions of $\beta$ for $\gamma=0.1$}.}
\label{Fugure 4}%
\end{center}
\begin{center}
\includegraphics[
height=2.0609in, width=3.1514in
]%
{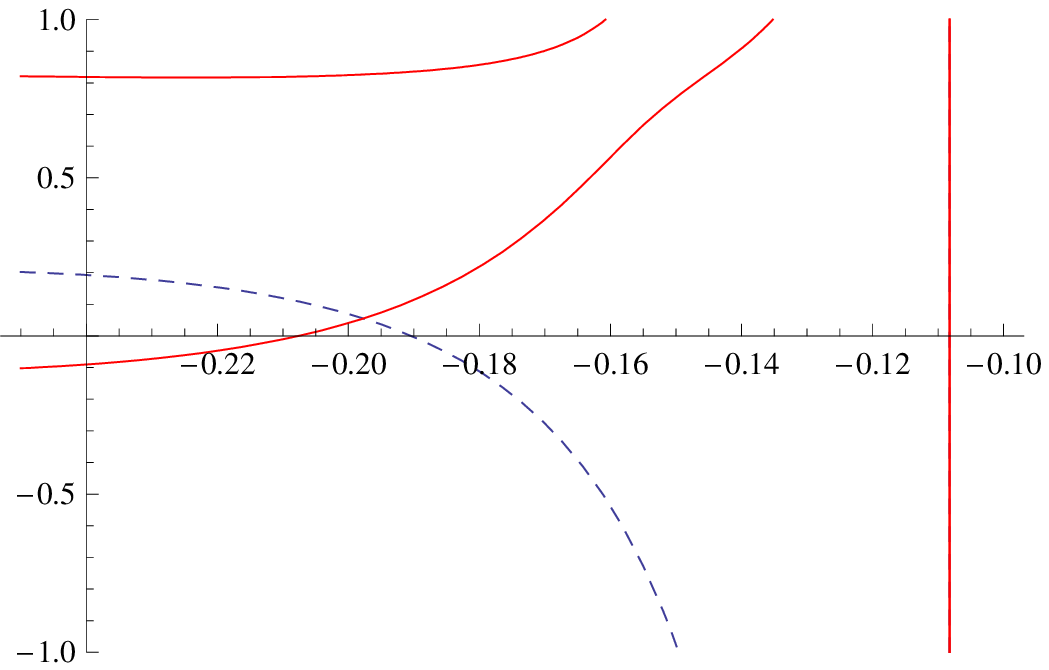}%
\\
{\small Figure 4: \emph{The two $10^{-1}\times$ mass$^{2}$
eigenvalues and $V$ (dashed)  as functions of $\beta$ for
$\gamma=0.1$.}}
\label{Figure 5}%
\end{center}
From figure 5 it can be seen that these solutions have a net
orientifold
charge.%
\begin{center}
\includegraphics[
height=2.0323in, width=3.1514in
]%
{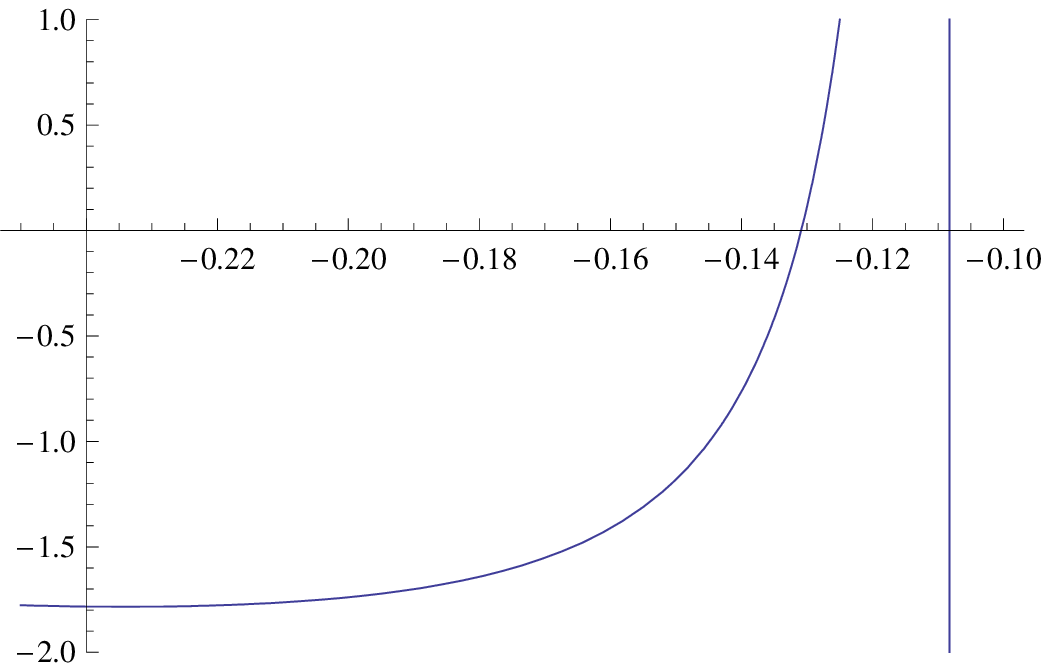}%
\\
{\small Figure 5: \emph{$V_{O6}$ as a function of $\beta$ for
$\gamma=0.1$.}}
\label{Figure 6}%
\end{center}


From figure 5 we can see that, if $\beta$ is chosen near $-0.13$ the
solution has a vanishing O6/D6 charge. At this value of $\beta$ both
the mass matrix eigenvalues are positive as can be seen from figure
4. So, we get AdS solutions with vanishing charge that are stable in
the $\rho,\tau$-directions.

\subsection*{The scales }
Note that in the previous we have absorbed $\rho$ and $\tau$ in the
various fluxes such that it did not appear explicitly in the
equations (\ref{potentials0}, \ref{potentials1}) and
(\ref{potentials2}). We can therefore choose them at will by
rescaling the various fluxes (and $\mu$). This implies that we can
make the solution as weakly coupled as we want and choose the volume
such that we can neglect $\alpha'$ corrections and perhaps still
have a decoupling of KK modes \cite{Caviezel:2008ik}.  However there
is a danger since we also have to rescale $\mu$, but, as we
explained before, the number of orientifolds is not a free
parameter. Furthermore, scaling of fluxes is also potentially
dangerous because of quantisation. So, it remains to be seen whether
a given explicit model fulfills the right conditions.

If we reinstate the dependence of $\rho$ and $\tau$ in the equations
we can plot the potential in function of $\rho$ and $\tau$. This we
have done in figure 6 for $\beta=-0.2$. We have chosen $\rho$ and
$\tau$ such that the dS minimum derived above is at $\rho=\tau=1$.
We can clearly see the minimum, and in addition an inflexion point
near $\left( \rho,\tau\right) =\left( 1.08,1.25\right)$ acting as a
barrier against a deeper drop in the potential towards the upper
right in the picture. This is qualitatively the same kind of
behavour as in KKLT \cite{Kachru:2003aw} and suggests a dS vacua
non-perturbatively unstable against tunneling to a lower energy.
However, one needs to be very careful when drawing these kinds of
conclusions. The only critical point in figure 6 that we have
actually proven to be a solution to the 10D equations of motion is
the minimum at $\left( \rho,\tau\right) =\left(  1,1\right)  $. Any
other critical points generated by moving off in the $\left(
\rho,\tau\right)  $-plane are likely not to be full solutions.
\begin{center}
\includegraphics[
height=3.1185in, width=3.1514in
]%
{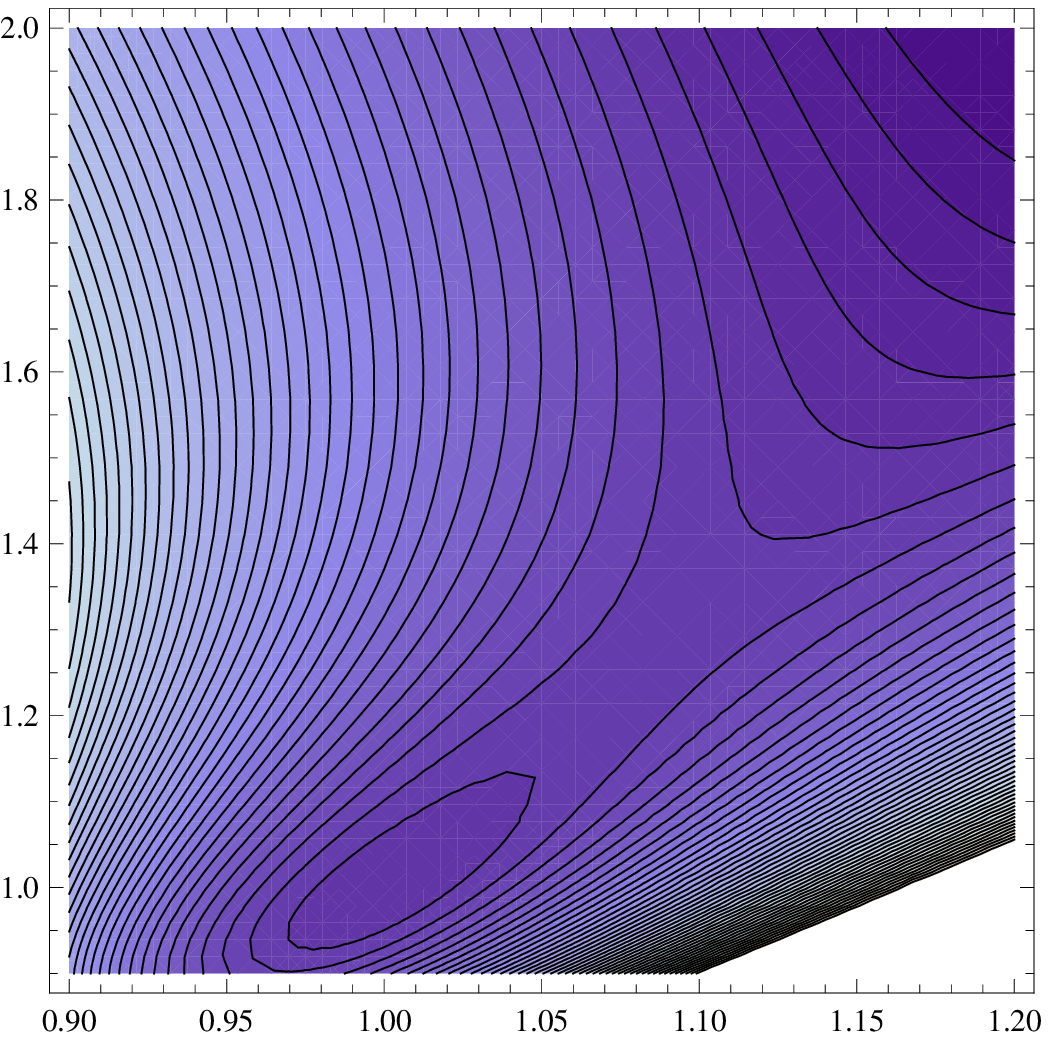}%
\\
{\small Figure 6: \emph{$V$ as a function of $\rho$ and $\tau$ for
$\beta=-0.2$ and $\gamma=0.1$.}}
\label{Figure 7}%
\end{center}

\subsection*{The coset geometries}
So far we have not realised explicit geometries for these torsion
classes. In total we have 2 conditions on the torsion classes.
\begin{enumerate}
\item $\d W_2\propto \Omega_R$\,,
\item $W^2_{ij}=\tfrac{1}{6}W_2^2g_{ij} +i\alpha(JW)_{ij}$\,,
\end{enumerate}
where the Einstein equations dictate that $\alpha$ is related to the
fluxes as follows
\begin{equation}\label{alpha2}
\frac{\alpha}{iW_1}=\Bigl(\frac{2\gamma-1}{2+8\gamma}\Bigr)\,\,\frac{8(3\gamma^2
-\beta^2)+(2\gamma-1)(1+4\gamma)}{(2\gamma-1)^2-\beta^2}\,.
\end{equation}
We will now check these conditions for the coset geometries and the
Iwasawa manifold discussed in \cite{Caviezel:2008ik} \footnote{Another set of explicit $\SU(3)$-
structure manifolds appeared in \cite{Popov:2009nx}. These spaces have $W_1=0$ and
$W_2\neq 0$. We have verified that this does not allow the dS solutions
we have considered.}

One finds that the only examples that can fulfill $\d W_2\sim
\Omega_R$ and the degeneracy condition  are $Sp(2)/S(U(2)\times
U(1))$ and $SU(3)/U(1)\times U(1)$. It turns out $Sp(2)/S(U(2)\times
U(1))$ is a special subcase of $SU(3)/U(1)\times U(1)$ when some
moduli are fixed; so we only discuss the coset $SU(3)/U(1)\times
U(1)$. According to \cite{House:2005yc,Caviezel:2008ik} we have
\begin{align}
J  & =-ae^{12}+be^{34}-ce^{56}\\
\Omega & =d\left(  \left(  e^{245}+e^{135}+e^{146}-e^{236}\right)
+i\left( e^{235}+e^{136}+e^{246}-e^{145}\right)  \right)\,,
\end{align}
where the $e^i$ are the Cartan--Maurer forms. The metric is diagonal
with respect to the Cartan--Maurer forms and is given by
\begin{equation}
g=\left(
\begin{array}
[c]{cccccc}%
a &  &  &  &  & \\
& a &  &  &  & \\
&  & b &  &  & \\
&  &  & b &  & \\
&  &  &  & c & \\
&  &  &  &  & c
\end{array}
\right)\,.
\end{equation}
For convenience we introduce the notation  $g=(a,b,c)$. We
furthermore have
\begin{align}
W_{1}& =\frac{i}{3}\,\frac{a+b+c}{\sqrt{abc}}\,,\\
W_{2}& =-\frac{2i}{3\sqrt{abc}}\left(  a\left(  2a-b-c\right)
e^{12}+b\left( a-2b+c\right)  e^{34}+c\left(  -a-b+2c\right)
e^{56}\right)\,.
\end{align}
From these expressions we find
\begin{align}
|W_{2}|^2  & =\frac{16}{3abc}\left(  a^{2} +b^{2}+c^{2}-\left(
ab+ac+bc\right)  \right)\,,\\
(W^2_2)_{nm}&=-\frac{4}{9abc}\left( a\left( 2a-b-c\right)
^{2},b\left( 2b-a-c\right)^{2},c\left( 2c-a-b\right)
^{2}\right)\\
(JW_2)_{mn}&=\frac{2i}{3\sqrt{abc}}\left(  a\left( 2a-b-c\right)
,b\left( 2b-a-c\right)  ,c\left(  2c-a-b\right) \right)\,.
\end{align}
In general $(JW_2)_{mn}$ and $(W^2_2)_{mn}-\frac{1}{6}g_{nm}
W_{2}^{2}$ are not parallel to each other, but at, e.g., $a=b$ we
find
\begin{equation}
\alpha=\frac{2(c-a)}{3a\sqrt{c}}
\end{equation}
So, we should look for solutions with this value for $\alpha$ and
\begin{equation}
|W_{1}|^{2} =\frac{\left(2a+c\right)^{2}%
}{9a^{2}c}\,,\qquad |W_{2}|^{2} =\frac{16}{3a^{2}c}\left( a-c\right)
^{2},
\end{equation}
where $a,c>0$.

We have been able to find such solutions corresponding to new,
non-supersymmetric AdS vacua. For instance, with $\gamma=0.1$ we
find two solutions, both stable in the $\rho$ and $\tau$ directions
(we take $m^2=1$):
\begin{equation}
 a\approx 1.355\,,\qquad c\approx 0.5889\,,\qquad \beta
\approx-0.129\,.
\end{equation}
This solution has net D-brane charge (as can be verified using the
plots). The other solution has net O6 charge
\begin{equation}
a\approx 1.7625\,,\qquad c\approx 0.7718\,, \qquad \beta\approx
0.126\,.
\end{equation}
Such non-supersymmetric AdS4 vacua will be studied in more detail in
\cite{Koerber09}.

We have not been able to find any dS solutions for this coset, in
agreement with the results of \cite{Caviezel:2008tf}.

\subsection*{The Iwasawa manifold}
There is one extra example discussed \cite{Caviezel:2008ik} that can
satisfy the degeneracy condition and $\d W_2\propto \Omega_R$. This
is the Iwasawa manifold. In Cartan--Maurer basis the metric is given
by
\begin{equation}
g=\left(1, 1, y^2 \right)\,.
\end{equation}
with $y$ some fixed number.  Furthermore
\begin{align}
&J= e^{12} + e^{34} -y^2 e^{56}\,,\\
&W_2=-\frac{4i y}{3}(e^{12} + e^{34}+ 2y^2e^{56})\,,\\
&W_1=-\frac{2i y}{3}\,.
\end{align}
From this we have
\begin{equation}
\alpha=\frac{4y}{3}\,,\qquad |W_1|^2=\frac{4y^2}{9}\,,\qquad
|W_2|^2=\frac{64 y^2}{3}\,.
\end{equation}
We have not been able to find other vacuum solutions apart from the
susy AdS ones. So for the Nilmanifold it is possible to have the
susy choice for the fluxes and, at the same time, have the
degeneracy in the tensors $JW_2$ and $W_2^2$.

\section{Discussion}

In this paper we have investigated on general grounds the conditions
for the existence of classical de Sitter solutions in string theory.
We also went further by analysing specific Type IIA O6
constructions. The simplest models, in which the fluxes are closed
and non-exact, generically have moduli directions for which the
potential has no stationary dS point (not even an unstable one).
However, using $\SU(3)$-structure solutions as a testbed for models
that have different kinds of fluxes, we were able to find a simple
set of conditions on the torsion classes in order for specific de Sitter
solutions to exist. We explicitly verified these conditions for the
coset geometries and found that these conditions could almost be
satisfied but not quite. This we take as an indication that our
conditions, though non-trivial, are not impossible to be realized.
For the coset geometries that had the almost correct form of the
torsion classes we were able to find new non-supersymmetric AdS
solutions.

Concerning stability we have only investigated the masses of the
$\rho$ and $\tau$ scalars. The general moduli structure of these
generalised Calabi Yau spaces is not understood and can only be
studied when there exist an explicit geometry realising our
conditions on the torsion classes. However for the new AdS solutions
we have found in the $SU(3)/U(1)\times U(1)$ coset construction, it should be
possible to use the effective theory developed in
\cite{Caviezel:2008ik} (for a consistent subset of the degrees of
freedom) to study the stability.

Once an explicit geometry for the dS solution satisfying our
conditions is found, it is important to study the charge (and flux)
quantisation since the O6 charge depends on the involutions present
in the explicit geometry.
While our analysis relies on the smearing of the orientifolds, they
should be understood, in a fully microscopic construction of our dS
solutions, as a localized source whose singularity admits a stringy
resolution. We consider it as an important avenue for further
research to understand the effect of the backreaction of the sources
defined in this microscopic manner.

While SUSY AdS can be argued to be quite generic, dS solutions to
the equations of motion require fluke alignment of various
contributions to the internal Einstein equations. It therefore seems
likely that dS solutions should be regarded as accidental from a
landscape point of view. If one, furthermore, requires perturbative
stability in all directions, it might become exceedingly difficult
to find actual examples (see e.g. \cite{GomezReino:2008bi}). While
our analysis has been purely perturbative, there is no reason to
expect that the difficulties would go away in a non-perturbative
setting. Unfortunately, the presently available methods do not allow
for a detailed analysis of the non-perturbative case.

Given a critical point there is really no reason to expect a minimum
along a particular direction in moduli space. The critical point
might as well be a maximum or an inflexion point, and one might
argue that the chances for a given critical point to be a minimum in
one direction is only around $1/2$. If the dimensionality of the
moduli space is $N$, then the fraction of critical dS points that
actually are minima is down by a factor $2^{-N}$. With $N$ of the
order of a few hundred, this reduction with respect to the total
number of critical dS points in the landscape can easily be of the
same order, or even exceed, the expected $10^{-120}$ from the
smallness of the observed cosmological constant. One can therefore
argue that the existence of a perturbatively stable dS vacua, is at
least as severe a finetuning as the size of the cosmological
constant itself. It is in fact far from obvious that there are any
candidate vacua left in the landscape at all. Hence, it is
reasonable to investigate whether perturbatively unstable dS
critical points can work from a phenomenological point of view
\cite{Kallosh:2002gg}.

Finally we like to mention some interesting directions for further
research. One obvious direction is to find explicit geometries that
satisfy our conditions on the torsion classes needed for our simple
de Sitter solutions. Reference \cite{Grana:2006kf} contains an
explicit classification of Solvmanifolds which we plan to
investigate. If an explicit geometry can be found one can study the
stability of the solutions and the effect of the charge and flux
quantisation. On the other hand, when one considers explicit
geometries one can also allow more general fluxes then the one we
considered (those given by $\Omega, J$ and $W_2$) as was done for
instance in \cite{Caviezel:2008tf}. In general this is a hard
problem, but can be done if one can systematically scan the scalar
potential in these IIA orientifold models (see e.g.
\cite{Caviezel:2008ik, Villadoro:2005cu, Derendinger:2004jn}) for
critical points. In some interesting cases (like for twisted tori),
the effective theory is $\mathcal{N}=4$ gauged supergravities
\cite{Roest:2009dq, Dall'Agata:2009gv} which facilitate a systematic
scanning for de Sitter critical points \cite{deRoo:2003rm,
deRoo:2002jf}.

\section*{Acknowledgements}

We would like to thank Bret Underwood and Paul Koerber for useful
explanations and discussions on this topic. We also thank  Fernando
Marchesano, Paul McGuirk, Timm Wrase, Maxim Zabzine, and Marco
Zagermann for useful discussions. Finally, we thank Johan Blaback
for still spotting typos in v4 of this paper. T.V.R. is supported by
the G\"{o}ran Gustafsson Foundation. U.D. is supported by the
Swedish Research Council (VR) and the G\"{o}ran Gustafsson
Foundation. The work of SSH and GS  was supported in part by NSF
CAREER Award No. PHY-0348093, DOE grant DE-FG-02-95ER40896, a
Research Innovation Award and a Cottrell Scholar Award from Research
Corporation, a Vilas Associate Award from the University of
Wisconsin, and a John Simon Guggenheim Memorial Foundation
Fellowship. GS would also like to acknowledge support from the
Ambrose Monell Foundation during his stay at the Institute for
Advanced Study, Princeton  and the hospitality of the Institute for
Advanced Study, Hong Kong University of Science and Technology
during the final stage of this work.

\newpage
\appendix

\section{Form conventions and useful formulae}
A $p$-form $A_p$ in components is given by
\begin{equation}
A_p = \frac{1}{p!} A_{\mu_1\ldots \mu_p} \d x^{\mu_1} \wedge \ldots
\wedge \d x^{\mu_p}\,.
\end{equation}
Forms obey the following algebra
\begin{equation}
A_p \wedge B_q = (-)^{pq} B_q \wedge A_p\,.
\end{equation}
The exterior derivative is defined via
\begin{equation}
\d A_p=\frac{1}{p!}\partial_{[\nu}A_{\mu_1\ldots \mu_p]}\d
x^{\nu}\wedge \d x^{\mu_1}\wedge \ldots \wedge \d x^{\mu_p}\,,
\end{equation}
and obeys the Leibniz rule
\begin{equation}
\d(A_p\wedge B_q)=\d A_p\wedge B_q + (-)^p A_p\wedge \d B_q\,.
\end{equation}
In $D$ dimensions we define the epsilon \emph{symbol}
$\varepsilon_{\mu_1\mu_2\ldots\mu_p}$ via
\begin{equation}
\varepsilon_{01\ldots D-1}=1\,,
\end{equation}
and it is antisymmetric in all indices
$\varepsilon_{[\mu_1\mu_2\ldots\mu_p]}=\varepsilon_{\mu_1\mu_2\ldots\mu_p}$.
From the epsilon symbol we define the epsilon \emph{tensor}
$\varepsilon_{\mu_1\mu_2\ldots\mu_p}$ via
\begin{equation}
\epsilon_{\mu_1\mu_2\ldots\mu_p}=\sqrt{|g|}\varepsilon_{\mu_1\mu_2\ldots\mu_p}\,.
\end{equation}
Contractions of the epsilon tensor (and symbol) obey the following
relations
\begin{equation}
\epsilon_{\mu_1\mu_2\ldots\mu_q\mu_{q+1}\ldots\mu_D}\epsilon^{\mu_1\mu_2\ldots\mu_q
\nu_{q+1}\ldots\nu_D}=(-)^tq!(D-q)!\,\,\delta^{[\nu_{q+1}}_{[\mu_{q+1}}\ldots
\delta^{\nu_D]}_{\mu_{D}]}\,,
\end{equation}
where $t$ stands for the number of timelike dimensions of the
$D$-dimensional space. The Hodge operator $\star$ maps $p$-forms
into $(D-p)$-forms. We define $\star$ on the coordinate $p$-forms
and by linearity it is defined on all forms
\begin{equation}
\star (\d x^{\mu_1}\wedge \ldots \wedge \d
x^{\mu_p})=\frac{1}{(D-p)!}\epsilon_{\nu_1\ldots\nu_{D-p}}^{\qquad\mu_1\ldots\mu_p}\,\d
x^{\nu_1}\wedge \ldots \wedge \d x^{\nu_{D-p}}\,.
\end{equation}
The $\star$ operation has the following properties
\begin{align}
&\star A_p\wedge
B_p=\star B_p\wedge A_p=\frac{1}{p!}A_{\mu_1\ldots\mu_p}B^{\mu_1\ldots\mu_p}\star 1\,,\\
&\star\star A_p=(-)^{p(D-p)+t}A_p\,.
\end{align}
Useful identities are
\begin{align}
&\d x^{\mu_1}\wedge\ldots\wedge \d
x^{\mu_D}=(-)^t\varepsilon^{\mu_1\ldots\mu_D}\d x^0\wedge
\ldots\wedge \d x^{D-1}\,,\\
&\star 1=\sqrt{|g|}\d x^0\wedge \ldots\wedge \d x^{D-1}\,.
\end{align}
As an application of these conventions one has
\begin{equation}
\star_{10}(A_p\wedge B_q)=(-1)^{p(6-q)}\star_4A_p\wedge \star_6
B_q\,,
\end{equation}
where $A$ is a form in four-dimensional spacetime and $B_q$ is a
form on the internal six-dimensional space.

For a metric of the form
\begin{equation}
\d s_{10}^2 = \tau^{-2}\e^{2\alpha A(y)}g^4_{\mu\nu}\d x^{\mu}\d
x^{\nu} +\rho\e^{2\beta A(y)}g^6_{ij}\d y^i\d y^j\,,
\end{equation}
the Ricci tensor is (assuming constant $\tau$ and $\rho$)
\begin{align}
\mathcal{R}^{10}_{\mu\nu}=&R_{\mu\nu}(g^4)-4(\alpha^2 +
\alpha\beta)\e^{2(\alpha-\beta)A(y)}(\partial
A)^2\tau^{-2}\rho^{-1}g^4_{\mu\nu}\,,\nonumber\\
&- \alpha\e^{2(\alpha-\beta)A}\tau^{-2}\rho^{-1}g^4_{\mu\nu}\Box A  \,,\\
\mathcal{R}^{10}_{i j}=&R_{ij}(g^6)-4(\beta^2 +
\alpha\beta)(\partial A)^2g^6_{ij} + 4(\beta^2-\alpha^2
+2\alpha\beta)\partial_i A
\partial_j A\nonumber
\\& -4(\alpha+\beta)\nabla_i\partial_j A -\beta g^6_{ij}\Box A \,.
\end{align}

\section{10D Einstein and dilaton equation }

The 10D action is (where we have put $\kappa^2_{10}$=1/2)
\begin{equation}
\int\sqrt{g}\Bigl\{R-\tfrac{1}{2}(\partial\phi)^2-\sum_n\frac{1}{2\,n!}\e^{a_n\phi}F_n^2\Bigr\}
+ S_{loc}\,,
\end{equation}
where $\Sigma_n$ represents the sum over all the field strengths and
the numbers $a_n$ are given by
\begin{equation}
a^{RR}_n=\frac{5-n}{2}\,,\qquad a^{NS}_3=-1\,.
\end{equation}
In IIA the RR field strengths are $F_0, F_2, F_4$. When
space-filling $F_4$ flux is considered we will define it using
$F_6$. In IIB the RR fields strengths are $F_1, F_3, F_5$, where
$F_5$ is assumed to be self-dual. The source action is
\begin{equation}
S_{loc}=-\int_{p+1}T_p\sqrt{|g|} +\mu_p\int_{p+1}C_{p+1}\,,\qquad
T_p=\pm |\mu_p|\e^{(p-3)\phi/4}\,.
\end{equation}
where the plus sign is for $D$-branes and the minus sign for
orientifold planes.

The Einstein equation is given by (for $\phi$ constant)
\begin{equation}
R_{ab}=\sum_n\bigl( -\frac{n-1}{16n!}g_{ab}\e^{a_n\phi}F_n^2 +
\frac{1}{2(n-1)!}\e^{a_n\phi}(F_n)^2_{ab} \bigr) +
\tfrac{1}{2}(T^{loc}_{ab}-\tfrac{1}{8}g_{ab}T^{loc})\,,
\end{equation}
where the local stress tensor reads
\begin{equation}
T_{\mu\nu}^{loc}=-T_p\,g_{\mu\nu}\,\delta(\Sigma)\,,\qquad
T_{ij}=-T_p\,\Pi_{ij}\,\delta(\Sigma)\,.
\end{equation}
Throughout $a,b$ are 10D indices, $i,j$ are internal and $\mu\nu$
are external. $\Pi_{ij}$ is the projector on the cycle wrapped by
the source. In the smeared limit (which is considered when $p>3$) we
have
\begin{equation}
\delta(\Sigma)\rightarrow 1\,,\qquad \Pi_{ij}\rightarrow
\frac{p-3}{6}\,\,g_{ij}\,.
\end{equation}
These equations, that define the smeared sources, are not always
that simple, but they are valid for the cases we study in this
paper. In general, there could be traceless contributions as well.
Taking the trace over the internal indices and integrating over the
6D space one finds ($V_p=T_p$):
\begin{equation}\label{condition1}
-V_R=\sum_n\frac{(n+3)}{4}V_n +\frac{1}{8}(15-p)V_p\,.
\end{equation}
The 10D dilaton equation is
\begin{equation}
\Box\phi=0=\sum_n
\frac{a_n}{2n!}\e^{a_n\phi}F_n^2\pm\frac{p-3}{4}\e^{(p-3)\phi/4}|\mu_p|\,\delta(\Sigma)\,,
\end{equation}
from which we have
\begin{equation}\label{condition2}
\sum_n a_nV_n +\frac{p-3}{4}V_p=0\,.
\end{equation}

From the expression for the effective potential we find:
\begin{align}
&\partial_{\rho}V=0:\qquad  -V_R  -3V_H + \sum_{q}(3-q)V_q + \frac{(p-6)}{2}V_{p}=0 \,,\\
& \partial_{\tau}V=0:\qquad -2V_R -2V_H - 4\sum_{q}V_q -3V_{p}=0\,,
\end{align}
where $q$ runs over the RR field strengths.  We notice that
(\ref{condition1}) can be found from summing $2/3$ times the first
equation with the second equation. Equation (\ref{condition2}) can
be obtained from summing $-2$ times the first equation with the
second equation.\\

The trace of the Einstein equation over the external indices just
sets the value of the cosmological constant. This can best be seen
using the ordinary Einstein equation
\begin{equation}
G_{\mu\nu}=\sum_n\frac{1}{n!2}\e^{a_n\phi}\bigl(n\,(F_n^2)_{\mu\nu}
- \frac{1}{2}g_{\mu\nu}F_n^2 \bigr) +
\frac{1}{2}T^{local}_{\mu\nu}\,.
\end{equation}
When we take indices in the 4D spacetime we have\footnote{in IIA
with space filling $F_4$ we replace the space-filling component by
$F_6$. In IIB with non-zero $F_5$ this term is non-zero but if one
defines $V_5$ with an extra factor of $1/4$ the expressions match.}
$(F_n^2)_{\mu\nu}=0$. When we take the trace over the 4D indices and
remember that using $R_{10}=R_4 + R_6$ and $R_4=2V$ we recover the
definition of V
\begin{equation}
V=V_R + \sum_n V_n + V_p\,.
\end{equation}

\section{IIA SUGRA}
The form equations of motion are
\begin{align}
& \d (\star\e^{3\phi/2} F_{2}) + \e^{\phi/2}\star F_{4}\wedge H=0\,,\label{F_2 eq}\\
& \d (\star \e^{\phi/2} F_{4})- F_{4}\wedge H=0\,,\label{F 4 eq}\\
& \d(\star\e^{-\phi}H)  + \e^{\phi/2}\star F_4\wedge F_2
-\tfrac{1}{2}F_4\wedge F_4 + F_0\e^{3 \phi/2}\star F_2=0\,,\label{H eq}\\
& \d\star\d\phi -\tfrac{1}{4}\e^{\phi/2}\star F_4\wedge F_4 +
\tfrac{1}{2}\e^{-\phi}\star H\wedge H -
\tfrac{3}{4}\e^{3\phi/2}\star F_2\wedge F_2 -
\tfrac{5}{4}\e^{5\phi/2}\star F_0\wedge F_0=0\,,\label{dilaton eq}
\end{align}
where $F_0$ is the Romans' mass. The Bianchi identities read
\begin{equation}\label{Bianchi}
\d H_3=0\,,\qquad \d F_2=F_0H\,,\qquad\d F_4=F_2\wedge H_3\,.
\end{equation}
The Einstein equation is given by
\begin{align}
&0=\mathcal{R}_{MN}-\tfrac{1}{2}\partial_{M}\phi\partial_{N}\phi-\tfrac{1}{12}\e^{\phi/2}F_{MPQR}F_N^{\,\,PQR}
+ \tfrac{1}{128}\e^{\phi/2}g_{MN}F_4^2 -
\tfrac{1}{4}\e^{-\phi}H_{MPQ}H_{N}^{\,\,PQ} \label{Einstein eq}\\& +
\tfrac{1}{48}\e^{-\phi}g_{MN}H^2
-\tfrac{1}{2}\e^{3\phi/2}F_{MP}F_N^{\,\,P}
+\tfrac{1}{32}\e^{3\phi/2}g_{MN}F_2^2-
\tfrac{1}{16}g_{MN}\e^{5\phi/2}F_0^2\,.\nonumber
\end{align}

\section{$SU(3)$-structure equations}
Fluxes in IIA SUGRA lead to $\SU(3)$-structures as can be derived
from the existence criterium of a everywhere non-vanishing spinor on
the internal manifold. Out of the spinor bilinears one can define a
real two form $J$ and an imaginary self-dual three form $\Omega$
\cite{Lust:2004ig}. These forms satisfy many relations and we list
those that are not presented in the main text and which are
necessary for the computations presented in this paper:
\begin{align}
& \star_6\Omega=-i\Omega\,,\qquad\star_6 J=\tfrac{1}{2}\,J\wedge J  \,,\\
& \Omega\wedge\Omega^*=\tfrac{4i}{3}\,J\wedge J\wedge J\,,\qquad
J\wedge J\wedge J= 6 \epsilon_6\,,\\
& \Omega\wedge J=0\,,\qquad W_2\wedge J\wedge J=0\,,\\
&W_2\wedge \Omega=0\,,\qquad
\star_6W_2=-J\wedge W_2\,, \\
& J_{mn}W_2^{mn}=0\,,\qquad
J_{m}^{\,\,\,n}J_p^{\,\,\,q}(W_2)_{nq}=(W_2)_{mp}\,,\\
& (\Omega_R)^2_{ab}=(\Omega_I)^2_{ab}=4g_{ab}\,,\qquad
J^2_{ab}=g_{ab}\,.
\end{align}

The notation we use for ``squaring'' a tensor $T_{i_1\ldots i_n}$ is
\begin{equation}
T^2_{ij}=T_{ii_2i_3\ldots i_n}T_j^{\,\,i_2i_3\ldots i_n}\,.
\end{equation}

\bibliography{groups}
\bibliographystyle{utphysmodb}
\end{document}